# Technical Report: The Need for a (Research) Sandstorm through the Privacy Sandbox

Yohan Beugin
University of Wisconsin–Madison
ybeugin@cs.wisc.edu

Patrick McDaniel
University of Wisconsin–Madison
mcdaniel@cs.wisc.edu

## Abstract

The Privacy Sandbox, launched in 2019, is a series of proposals from Google to reduce *‘‘cross-site and cross-app tracking while helping to keep online content and services free for all’‘*. Over the years, Google implemented, experimented, and deprecated some of these APIs into their own products (Chrome, Android, etc.) which raised concerns about the potential of these mechanisms to fundamentally disrupt the advertising, mobile, and web ecosystems. As a result, it is paramount for researchers to understand the consequences that these new technologies, and future ones, will have on billions of users if and when deployed. In this report, we outline our call for privacy, security, usability, and utility evaluations of these APIs, our efforts materialized through the creation and operation of **Privacy Sandstorm**; a research portal to systematically gather resources (overview, analyses, artifacts, etc.) about such proposals. We find that our inventory provides a better visibility and broader perspective on the research findings in that space than what Google lets show through official channels.

**Note:** While Google deprecated in October 2025 most of the Privacy Sandbox APIs, we argue that it is still important to evaluate these and other proposals; if only to create more secure and private ones in the future.

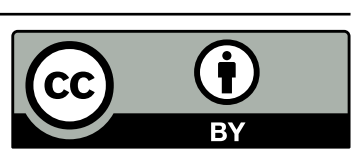

## 1 Introduction

The Privacy Sandbox initiative was launched by Google in 2019 to reduce cross-site and cross-app tracking. As part of this project, Google proposed several alternative mechanisms to restrict the use of third-party cookies, fingerprinting techniques, and other identifiers while still supporting key use cases such as advertising, bot and spam protection, conversion measurement, etc.

Over the past few years, Google has implemented, shipped, and deprecated some of these new APIs and technologies directly into their products (Chrome, Android, etc.). Google often did so without reaching consensus with other web browsers (e.g., Brave, Mozilla, Safari, etc.) on standardization and on these new proposals' privacy claims. This raised several concerns about the potential of these mechanisms put forward by Google to fundamentally disrupt the advertising, mobile, and web ecosystems. As a result, we believe that it is important to transcend the debate around the somewhat controversial nature of Google's project to ask the following questions: *how can we contribute as a research community to the discussions about the Privacy Sandbox (and similar proposals put forward by others) and have real impact to protect the privacy of billions of users?* Indeed, it appears paramount for researchers to understand the consequences that these new technologies (and future ones) will have on real users if and when deployed.

In the following report, we outline our call for privacy, security, usability, and utility evaluations of these APIs. Concretely, our efforts materialized through the creation and operation of **Privacy Sandstorm**; a research portal presenting a roadmap to these rapidly evolving changes, systematically gathering and publishing resources (overview, analyses, artifacts, etc.) about such proposals, and giving broader visibility to the research findings in that space. The Privacy Sandstorm website was launched during our HotPETS 2024 talk [9] with the goal to spark and coordinate a conversation among attendees on the steps that we should take, as a research community, to better understand these changes and play a part in them.

As we systematically list and provide visibility to all research findings in that space, we find that we offer a broader perspective on the findings related to the Privacy Sandbox than what Google lets show through their official channels. We expect our resources to not only be useful to the academic and industry communities, but also to en-courage people to contribute to them (open to suggestions, either by message or directly through opening a pull request on the repository) and join our endeavor (see invite link to communication workspace on website).

## 2 Privacy Sandstorm: a Research Portal

Privacy Sandstorm is a research portal that gathers resources about the Privacy Sandbox initiative from Google as well as other proposals related to online security, privacy, and advertising. Our main objectives are two-fold:

1. Giving broader visibility to the findings from the research community in that space.
2. Coordinating multidisciplinary approaches to evaluate and improve these proposals.

To do so, we continuously perform an inventory of the relevant analyses that were carried out for each proposal from the Privacy Sandbox and similar mechanisms, and we gather different resources such as datasets and software that could be useful to researchers to analyze the claims made by these proposals. This inventory can be found on our website **Privacy Sandstorm**, as well as in this technical report whose content is automatically generated from the content of the website. For each identified API and proposal, we provide an overview of its goals, how it works, and link to the official explainer and documentation. We also list analyses and evaluations that have been performed by several actors along with links and pointers to their corresponding resources (papers, artifacts, videos, slides, etc.).

## 3 Datasets & Software

Our objective here is to list different resources that could be useful to researchers who are evaluating the privacy claims of different advertising and web proposals. Check out also similar content that is located elsewhere on our website under the “Datasets & Software” tag.

### 3.1 Datasets

- **HTTP Archive**: This project regularly crawls top websites to record different information about the resources being fetched, APIs used, etc. The dataset and historical crawls can be accessed through BigQuery.
- **HTTP Archive - Topics Classification**: We classified all unique hostnames from all requests present in the HTTP Archive dataset that were made between November 2010 and June 2024 with the latest version of the Topics API classifier (i.e., `chrome5`). The classification spans a total of more than 147 million hostnames corresponding to 31 million unique domains. You can access the results either through the new `GET_HOST_CATEGORIES` BigQuery function on the HTTP Archive dataset or by directly querying the raw data in the `httparchive.urls.categories` table.
- **.well-known dataset**: attestations and related website sets discovered by the corresponding scanner are made publicly available.
- **Criteo Research Datasets**: Criteo, a French-based advertising company, has released over the years several datasets related to advertising campaigns, clicks, conversion, etc.
- **A web tracking data set of online browsing behavior of 2,148 users**: this is an anonymized dataset collected in October 2018 on 2148 users from Germany who have volunteered to share their desktop browsing histories for a financial compensation.
- **National Internet Observatory**: the National Internet Observatory is a research study that collects data on the online behaviors and habits of U.S. participants (computers, phones, tablets). The Observatory aims to help researchers understand how people behave online and how online platforms structure what people see. Researchers can apply for data access by submitting proposals for studies.
- **Fingerprinting Study**: dataset of browser attributes with users’ demographics details for 8,400 U.S. participants.
- **Carat: Collaborative Energy Diagnosis**: the Carat research team from University of Helsinki led a project collecting large-scale energy diagnostics (e.g., battery status) from smartphone applications and devices at regular intervals. The dataset, if still available, could potentially be useful to analyze user activities on mobile devices and applications usage (Topics for Android, etc.)
- **DuckDuckGo Tracker Radar**: dataset of most common third-party domains found on the web.

### 3.2 Software

- **Topics Classifier**: this repository reproduces Google’s implementations of the Topics API for the Web and for Android.
- **.well-known scanner**: we used to regularly scan the Web for the presence of `.well-known` resources and files that were introduced by the Related Website Sets and Attestation from the Privacy Sandbox.
- **VisibleV8**: this is a custom variant of the V8 JavaScript engine at the heart of Chromium, the Node.js runtime, etc., that allows to capture and log JavaScript API calls while crawling websites for instance. This could be useful to researchers performing measurement studies on the Web.
- **DuckDuckGo Tracker Radar Collector**: a modular, multithreaded, puppeteer-based crawler used to generate third party request data for DuckDuckGo Tracker Radar, but also used by several academic studies.

### 3.3 Other

- **SoK: Advances and Open Problems in Web Tracking** for a systematization of developments in web tracking (threat model for the web, tracking techniques, defenses, regulations overview, and evolution) and identification of open problems in the field.

## 4 Privacy Sandbox Proposals

**October 2025**: Google officially announced the deprecation of (most of) the Privacy Sandbox APIs for the Web and Android, see this status overview.

The Privacy Sandbox is an initiative from Google to reduce *"cross-site and cross-app tracking while helping to keep online content and services free for all"*. It is composed of more than 20 proposals that can fundamentally disrupt the advertising, mobile, and web ecosystems.

The Privacy Sandbox for the Web is a series of proposals made by Google to phase out third-party cookies and reduce web tracking while still supporting cross-site use cases such as ads targeting or robot and spam detection. To do so, Google proposes to replace some of the current web components and APIs, that have been diverted from their initial purpose to enable extensive tracking (e.g., third party cookies), with privacy-preserving alternatives.

The goals of the Privacy Sandbox on Android are similar; reduce user tracking on Android by deprecated access to cross-app identifiers such as Advertising ID and limiting the scope that third party libraries in smartphone applications can access.

You will find below a list of all proposals in the Privacy Sandbox mapped to corresponding analyses from different actors.

### 4.1 General Analyses

We list here general analyses about the Privacy Sandbox broadly; more specific analyses can be found next under each individual proposal.

#### 4.1.1 API Adoption

You can find next a list of general measurement studies and analyses of the adoption of several APIs from the Privacy Sandbox. More specific measurements of individual APIs are listed directly under each corresponding proposal.

- [37] Garrett A Johnson and Nico Neumann. 2024. The Advent of Privacy-Centric Digital Advertising: Tracing Privacy-Enhancing Technology Adoption. (2024). Retrieved from https://pep.gmu.edu/wp-content/uploads/sites/28/2024/04/Johnson-Neumann.pdf
- [38] Garrett Johnson. 2024. Unearthing Privacy-Enhancing Ad Technologies (PEAT): The Adoption of Google's Privacy Sandbox. https://doi.org/10.2139/ssrn.4983927
- [11] Yohan Beugin, Paul Barford, and Patrick McDaniel. 2026. Lessons from the Adoption and Deprecation of the Privacy Sandbox Web APIs. https://doi.org/10.48550/arXiv.2606.26390
- [27] Rachid Youssef Grib, Alberto Verna, Nikhil Jha, Martino Trevisan, and Marco Mellia. 2026. The Rise and Fall of Google's Privacy Sandbox. https://doi.org/10.48550/arXiv.2607.00693

#### 4.1.2 Other

We list below general analyses that do not correspond to a specific API, but that are about the Privacy Sandbox more broadly.

- [18] Bennett Cyphers. 2019. Don't Play in Google's Privacy Sandbox. Retrieved from https://www.eff.org/deeplinks/2019/08/dont-play-googles-privacy-sandbox-1
- [41] Michael Kleber. 2019. Privacy Model for the Web. Retrieved from https://github.com/michaelkleber/privacy-model
- [66] Peter Snyder. 2022. Privacy And Competition Concerns with Google's Privacy Sandbox. Retrieved from https://brave.com/web-standards-at-brave/6-privacy-sandbox-concerns/
- [23] French Center of Expertise for Digital Platorm Regulation (PEReN). 2022. Shedding Light On...The Privacy Sandbox: A Collection of Tools for Third-Party Cookieless Online Advertising. Retrieved from https://www.peren.gouv.fr/en/actualites/2022-04-28/_eclairage/_sur/_privacy/_sandbox/
- [26] IAB Tech Lab Privacy Sandbox Task Force. 2024. Privacy Sandbox Fit Gap Analysis for Digital Advertising. Retrieved from https://iabtechlab.com/standards/privacysandbox/
- [42] Shunto Kobayashi, Garrett Johnson, and Zhengrong Gu. 2024. Privacy-Enhanced versus Traditional Retarget-

ing: Ad Effectiveness in an Industry-Wide Field Experiment. https://doi.org/10.2139/ssrn.4972368
- [28] Zhengrong Gu, Garrett A. Johnson, and Shunto J. Kobayashi. 2026. Can Privacy Technologies Replace Cookies? Ad Revenue in a Field Experiment. *Proceedings of the National Academy of Sciences* 123, (May 2026), e2603752123. https://doi.org/10.1073/pnas.2603752123
- [6] Abir Benzaamia and Oana Goga. 2026. Privacy Settings and Ad Perception: The Shift from Third-Party Cookies to the Privacy Sandbox. In *Proceedings of the 2026 CHI Conference on Human Factors in Computing Systems (CHI '26)*, April 2026. Association for Computing Machinery, New York, NY, USA, 1–17. https://doi.org/10.1145/3772318.3790444

#### 4.1.3 Regulation Concerns

Below, you will find analyses and comments about regulation, compliance, and market competition concerns related to the Privacy Sandbox.

- [50] Mark Nottingham. 2021. Playing Fair in the Privacy Sandbox: Competition, Privacy and Interoperability Standards. https://doi.org/10.2139/ssrn.3891335
- [52] Lukasz Olejnik. 2023. On the Governance of Privacy-Preserving Systems for the Web: Should Privacy Sandbox Be Governed?. In *Handbook on the Politics and Governance of Big Data and Artificial Intelligence*, Andrej Zwitter and Oskar Gstrein (eds.). Edward Elgar Publishing, 279–314. https://doi.org/10.4337/9781800887374.00022
- [51] Lukasz Olejnik. 2023. Reconciling Privacy Sandbox Initiatives with EU Data Protection Laws. (2023). Retrieved from https://lukaszolejnik.com/stuff/PrivacySandbox_PAAPI_LLM_LO.pdf
- [48] Shaoor Munir, Konrad Kollnig, Anastasia Shuba, and Zubair Shafiq. 2024. Google's Chrome Antitrust Paradox. Retrieved December 2, 2025 from https://papers.ssrn.com/abstract=4780718

### 4.2 Attestation/Enrollment

Similar to the APIs, the attestation mechanism is likely being deprecated, see the official announcement from Google.

#### 4.2.1 Overview

In order to use some of the Privacy Sandbox APIs, API callers have to go through an enrollment process to declare that they will not abuse these APIs for cross-site re-identification, but only for their intended use cases. The legal implications of this commitment if not respected is quite unclear, but this allows these callers to obtain an attestation file that must be placed at the `.well-known` URI `/.well-know/privacy-sandbox-attestations.json` on the domain they registered to call these APIs from.

#### 4.2.2 How does it work?

Chrome ships with a preloaded file containing a list of domains that have an attestation file registered and should be allowed to call the Privacy Sandbox APIs requiring attestation: Attribution Reporting, Protected App Signals (Android only), Private Aggregation (Chrome only), Protected Audience, Shared Storage (Chrome only), and Topics.

- Enrollment

#### 4.2.3 Analyses

- [12] Yohan Beugin, Sam Dutton, Yana Dimova, Rowan Merewood, and Barry Pollard. 2024. The 2024 Web Almanac: Cookies. In *The 2024 Web Almanac*. HTTP Archive. https://doi.org/10.5281/zenodo.14065903

### 4.3 Attribution Reporting API

This API is deprecated, although Google said they would continue work on a similar proposal through the web standards process, see the official announcement and this status overview from Google.

#### 4.3.1 Overview

Ad conversion measurement often relies on third-party cookies. Browsers are restricting access to third-party cookies because these can be used to track users across sites and hinder user privacy. The Attribution Reporting API enables those measurements in a privacy-preserving way, without third-party cookies. This API enables advertisers and ad tech providers to measure conversions in the following cases:

- Ad clicks and views.
- Ads in a third-party iframe, such as ads on a publisher site that uses a third-party ad tech provider.
- Ads in a first-party context, such as ads on a social network or a search engine results page, or a publisher serving their own ads.

#### 4.3.2 How does it work?

The developer guide provides a list of use cases which covers: (a) **Event-level reports** associate a particular event on the ad side (a click, view or touch) with coarse conversion data. To preserve user privacy, conversion-side data is coarse, and noise is added to the reports that are not sent immediately. The number of conversions is also limited. (b) **Summary reports** provide a mechanism for rich metadata to be reported in aggregate, to better support use-cases such as campaign-level performance reporting or conversion values.

- Documentation (Web)
- Explainer
- Documentation (Android)

#### 4.3.3 Analyses

- [43] Elena Bakos Lang, Giacomo Pope, Giovanni De Ferrari, Huy Nguyen, Lydia Yao, Thomas Pornin, Tyler

Colgan, and Viktor Gazdag. 2024. Privacy Sandbox Aggregation Service and Coordinator. (April 2024). Retrieved from https://www.fox-it.com/media/m3yogjsq/_ncc_group_google_privacy_sandbox_public_report_v2.pdf
- [1] Hidayet Aksu, Badih Ghazi, Pritish Kamath, Ravi Kumar, Pasin Manurangsi, Adam Sealfon, and Avinash V. Varadarajan. 2024. Summary Reports Optimization in the Privacy Sandbox Attribution Reporting API. *Proceedings on Privacy Enhancing Technologies* (2024). Retrieved from https://petsymposium.org/popets/2024/popets-2024-0132.php
- [21] John Delaney, Badih Ghazi, Charlie Harrison, Christina Ilvento, Ravi Kumar, Pasin Manurangsi, Martin Pál, Karthik Prabhakar, and Mariana Raykova. 2024. Differentially Private Ad Conversion Measurement. *Proceedings on Privacy Enhancing Technologies* (2024). Retrieved from https://petsymposium.org/popets/2024/popets-2024-0044.php
- [82] Yingtai Xiao, Jian Du, Shikun Zhang, Wanrong Zhang, Qiang Yan, Danfeng Zhang, and Daniel Kifer. 2025. Click Without Compromise: Online Advertising Measurement via Per User Differential Privacy. https://doi.org/10.48550/arXiv.2406.02463

### 4.4 Bounce Tracking Mitigations

This API has been launched by default in Chrome for users who have opted-in to blocking third-party cookies.

#### 4.4.1 Overview

Mitigations against bounce tracking.
- Documentation
- Explainer

### 4.5 CHIPS

This API is still being maintained by Google in Chrome, even after the deprecation announcement of most other Privacy Sandbox APIs.

#### 4.5.1 Overview

Cookies Having Independent Partitioned State (CHIPS) allow web developers to specify that they would like the cookies that they are setting to be saved in a partitioned storage, i.e., in a separate cookie jar per top-level site. Cookies that have the `Partitioned` attribute set can only be accessed by the same third party and from the same top-level site where they were created in the first place. In other words, partitioned cookies can not be used for third-party tracking across websites and allow for the legitimate use of third-party cookies on a top-level site.

#### 4.5.2 How does it work?

Partitioned cookies are stored by compatible browsers using partitioned storage, i.e., the cookie is stored using two keys: the host key and a new partition key. This *partition key* is based on the scheme and eTLD+1 of the top-level URL of the visited site on which the request to set a partitioned cookie was made.
- **API:**
  - `Partitioned` attribute for the `Set-Cookie` HTTP header
  - Partitioned cookies must be set with `Secure` and are only set and sent over secure protocols. It is also recommended setting the `__Host` prefix on partitioned cookies.
- Documentation
- MDN Documentation
- Explainer

#### 4.5.3 Analyses

- [12] Yohan Beugin, Sam Dutton, Yana Dimova, Rowan Merewood, and Barry Pollard. 2024. The 2024 Web Almanac: Cookies. In *The 2024 Web Almanac*. HTTP Archive. https://doi.org/10.5281/zenodo.14065903
- [83] Maximilian Zöllner, Anja Feldmann, and Ha Dao. 2025. A First Look at~Cookies Having Independent Partitioned State. In *Passive and Active Measurement*, 2025. Springer Nature Switzerland, Cham, 182–196. https://doi.org/10.1007/978-3-031-85960-1_8
- [13] Yohan Beugin, Jannis Rautenstrauch, Martina Kraus, Chris Böttger, Brian Smith, and Barry Pollard. 2026. The 2025 Web Almanac: Cookies. In *The 2025 Web Almanac*. HTTP Archive. https://doi.org/10.5281/zenodo.18246755

### 4.6 Federated Credential Management API

This API is still being maintained by Google in Chrome, even after the deprecation announcement of most other Privacy Sandbox APIs.

#### 4.6.1 Overview

The Federated Credential Management (FedCM) API allows identity providers to build an SSO-login infrastructure facilitated by compatible web browsers that does not require the use of third-party cookies or redirects.

#### 4.6.2 How does it work?

A relying party (RP) can allow users to sign-in using their credentials and account on a trusted third-party identity provider (IdP) by using the FedCM API. The browser mediates requests and information exchange between the RP and IdP through the FedCM API by: - Gathering user content to login to the RP with the IdP. - Recording if a relationship has been established between specific IdPs and RPs. - Providing features to the IdP and RP specific to the user consent provided.
- **API:**
  - JS:
    - `navigator.credentials.get()`
    - `IdentityCredential.disconnect()`
    - `navigator.login.setStatus("logged-in"/”logged-out”)`

- Permissions Policy on iframe: `allow="identity-credentials-get"`
- Documentation
- MDN Documentation
- Explainer

#### 4.6.3 Analyses

- [80] Maximilian Westers, Andreas Mayer, and Louis Jannett. 2024. Single Sign-On Privacy: We Still Know What You Did Last Summer. In *2024 Annual Computer Security Applications Conference (ACSAC)*, December 2024. IEEE, Honolulu, HI, USA, 321–335. https://doi.org/10.1109/ACSAC63791.2024.00039
- [76] Šimon Vacek. 2024. FedCM API Integration into Keycloak. Bachelor Thesis. Retrieved from https://theses.cz/id/qldogj/

### 4.7 Fenced Frames

This API is still being maintained by Google in Chrome, even after the deprecation announcement of most other Privacy Sandbox APIs.

#### 4.7.1 Overview

Fenced frames are similar to iframes, i.e., embedded frames with HTML content, with the distinction that they enforce a strict boundary between the embedded content and the embedding page that can not access each other context's DOM. These fenced frames are intended to be used by other Privacy Sandbox APIs, such as the Protected Audience or Shared Storage APIs to prevent for instance publishers to learn any information about an ad auction winner and the ad displayed to users.

#### 4.7.2 How does it work?

Compatible web browsers serve fenced frames and manage the boundary with the embedding context by mapping the URL of the fenced frame to an opaque URL and restricting access to the resources available from inside a fenced frame.

- **API:**
  - JS:
    - `window.HTMLFencedFrameElement`
    - `window.FencedFrameConfig`
    - `FencedFrameConfig.setSharedStorageContext()`
  - HTTP headers:
    - `Sec-Fetch-Dest: fencedframe`
    - `Supports-Loading-Mode: fenced-frame`
- Documentation
- Explainer

### 4.8 FLoC API

This API is deprecated.

#### 4.8.1 Overview

Federated Learning of Cohorts (FLoC) was a proposal from Google to deprecate *third-party cookies* and *reduce cross-site tracking* while still *enabling interest-based advertising*. Google claimed that users could not be fingerprinted through FLoC because "thousands of users" would share the same cohort (*k-anonymity*).

#### 4.8.2 How does it work?

With FLoC, every week web browsers assign users to an interest group (or cohort) by computing a locality-sensitive hash (LSH) on users' browsing histories. The computed cohort is reported to a central server (controlled by Google for Chrome's initial implementation) that enforces k-anonymity by either (a) checking that the corresponding cohort is composed of enough users, or (b) merging the cohort with other cohorts until this is satisfied. Advertisers embedded on websites can observe the cohort IDs of the visiting users by calling the FLoC API.

- **API call:** `document.interestCohort()`
- Documentation
- Explainer

From our PETS'24 paper:

``*Independent analyses of FLoC revealed a variety of privacy concerns: (1) requirement in trusting a single actor to maintain adequate k-anonymity, (2) concern that cohort IDs could create or be linked to fingerprinting techniques, (3) risk of re-identifying users by tracking their cohort IDs over time and by isolating them into specific cohorts through Sybil attacks. Finally, while some parameters and details of FLoC were still unclear, advertisers also had concerns about how to interpret the cohort ID for utility. Google eventually dropped FLoC for the Topics API.*''

#### 4.8.3 Analyses

- [55] Deepak Ravichandran and Sergei Vassilvitskii. 2020. Evaluation of Cohort Algorithms for the FLoC API. (2020). Retrieved from https://raw.githubusercontent.com/google/ads-privacy/master/proposals/FLoC/FLOC-Whitepaper-Google.pdf
- [61] Peter Snyder and Brendan Eich. 2021. Why Brave Disables FLoC. Retrieved from https://brave.com/blog/why-brave-disables-floc/
- [19] Bennett Cyphers. 2021. Google's FLoC Is a Terrible Idea. Retrieved from https://www.eff.org/deeplinks/2021/03/googles-floc-terrible-idea
- [25] Alessandro Epasto, Andrés Muñoz Medina, Steven Avery, Yijian Bai, Robert Busa-Fekete, \relax CJ Carey, Ya Gao, David Guthrie, Subham Ghosh, James Ioannidis, Junyi Jiao, Jakub Lacki, Jason Lee, Arne Mauser, Brian Milch, Vahab Mirrokni, Deepak Ravichandran, Wei Shi, Max Spero, Yunting Sun, Umar Syed, Sergei Vassilvitskii, and Shuo Wang. 2021. Clustering for Private Interest-Based Advertising. In *Proceedings of the 27th ACM*

*SIGKDD Conference on Knowledge Discovery & Data Mining (KDD '21)*, August 2021. Association for Computing Machinery, New York, NY, USA, 2802–2810. https://doi.org/10.1145/3447548.3467180
- [56] Eric Rescorla and Martin Thomson. 2021. Technical Comments on FLoC Privacy. (June 2021). Retrieved from https://mozilla.github.io/ppa-docs/floc_report.pdf
- [7] Alex Berke and Dan Calacci. 2022. Privacy Limitations of Interest-Based Advertising on The Web: A Post-Mortem Empirical Analysis of Google's FLoC. In *Proceedings of the 2022 ACM SIGSAC Conference on Computer and Communications Security (CCS '22)*, November 2022. Association for Computing Machinery, New York, NY, USA, 337–349. https://doi.org/10.1145/3548606.3560626
- [39] Guillaume Kessibi, Aymen Ould Hamouda, Charly Poirier, and Antoine Boutet. 2022. A Complementary Utility and Privacy Trade-off Evaluation of Google's FloC API. (2022). Retrieved from https://inria.hal.science/hal-03953308v1/document
- [75] Florian Turati, Karel Kubicek, Carlos Cotrini, and David Basin. 2023. Locality-Sensitive Hashing Does Not Guarantee Privacy! Attacks on Google's FLoC and the MinHash Hierarchy System. *Proceedings on Privacy Enhancing Technologies* (2023). Retrieved from https://petsymposium.org/popets/2023/popets-2023-0101.php

### 4.9 IP Protection

This API is deprecated, see the official announcement and this status overview from Google.

#### 4.9.1 Overview

IP Protection is a proposal to relay requests originating from third-party context through a proxy to hide the user's IP address. Google was envisioning deploying it only in Chrome's Incognito mode before backing out altogether.
- Documentation
- Explainer

### 4.10 Privacy Budget

This API is deprecated.

#### 4.10.1 Overview

The Privacy Budget proposal aimed to reduce the amount of identifying information that a site or party would have been able to obtain on users in order to prevent them from being uniquely re-identifiable.
- Documentation
- Explainer

#### 4.10.2 Analyses

- [63] Peter Snyder and Ben Livshits. 2019. Brave, Fingerprinting, and Privacy Budgets. Retrieved from https://brave.com/web-standards-at-brave/2-privacy-budgets/
- [57] Eric Rescorla. 2021. Technical Comments on Privacy Budget. (2021). Retrieved from https://mozilla.github.io/ppa-docs/privacy-budget.pdf
- [5] Enrico Bacis, Igor Bilogrevic, Robert Busa-Fekete, Asanka Herath, Antonio Sartori, and Umar Syed. 2024. Assessing Web Fingerprinting Risk. https://doi.org/10.48550/arXiv.2403.15607

### 4.11 Private Aggregation API

This API is deprecated, see the official announcement and this status overview from Google.

#### 4.11.1 Overview

The Private Aggregation API is similar to the Attribution Reporting API, but is meant to measure more general single and cross-site events, supporting other Privacy Sandbox APIs, such as Shared Storage or Protected Audience.
- Documentation (Android)
- Explainer

### 4.12 Private State Tokens

This API is still being maintained by Google in Chrome, even after the deprecation announcement of most other Privacy Sandbox APIs.

#### 4.12.1 Overview

Private State Tokens (previously named Trust Tokens) enable trust in a user's authenticity to be conveyed in a cross-site manner in order to combat fraud and bots without the use of third-party cookies. The API conveys information about the user in a cross-context manner. For example, if the user solves a CAPTCHA on one site, information that the user can be trusted is then communicated to other sites that the user visits. The tokens are securely stored in the user's browser and can be used elsewhere to confirm the user's authenticity without revealing their identity. Trust established on one platform, like a social media site or email service, can be extended to other websites, such as publishers or online stores, without compromising user privacy or linking identities across platforms.

#### 4.12.2 How does it work?

- An **issuer** website verifies that a user meets some definition of trustworthiness through a challenge; e.g., solving a CAPTCHA, logging into an account, performing a transaction, etc. If completed successfully, the user's web browser is issued a token that proves cryptographically that they are considered trustworthy by the issuer service.
- On the **redeemer** side, the website can then confirm that a user is considered to be real by an issuer the redeemer trusts by checking if they have a valid and recently issued token. If so, they can redeem these tokens as needed without having to prompt users for another challenge.
- **API calls:**

- Check if trust token exists: `document.hasPrivateToken()`
    - Check if redemption record exists: `document.hasRedemptionRecord()`
    - HTTP headers: `Sec-Private-State-Token`, `Sec-Private-State-Token-Lifetime`, `Sec-Private-State-Token-Version`, `Sec-Redemption-Record`
    - Issuance and redemption can be performed with Fetch requests and by embedding iframes with the `privateToken` attribute.
    - Additionally, the mechanism includes multiple new "Sec-" HTTP headers that the browser includes automatically in outgoing network requests.
    - The issuance and redemption server needs to set up additional encryption and guarantee mechanisms to ensure that tokens are in compliance with the mechanism requirements.
- Documentation
- Explainer
- Chrome Design Doc

#### 4.12.3 Analyses

- [2] Mir Masood Ali, Binoy Chitale, Mohammad Ghasemisharif, Chris Kanich, Nick Nikiforakis, and Jason Polakis. 2023. Navigating Murky Waters: Automated Browser Feature Testing for Uncovering Tracking Vectors. In *Proceedings 2023 Network and Distributed System Security Symposium*, 2023. Internet Society, San Diego, CA, USA. https://doi.org/10.14722/ndss.2023.24072

### 4.13 Protected Audience API (FLEDGE)

This API is deprecated, see the official announcement and this status overview from Google.

#### 4.13.1 Overview

The objective of the Protected Audience API (previously named FLEDGE) is to enable ``*On-device ad auctions to serve remarketing and custom audiences, without cross-site third-party tracking*''. The API provides a real-time bidding infrastructure that instead of using third-party cookies rely on a new storage mechanism in the browser called interest groups. Advertisers can use these interest groups to save and record in users' browser their interests (metadata) and ads to show later to users. This information can be used in future ad auctions by advertisers to bid on available interest groups to serve targeted ads in specific ad campaigns.

#### 4.13.2 How does it work?

When users visit websites and interact with specific products for instance, embedded advertisers on that site can record this by requesting browsers to add these users to a specific interest group that contains ads for the corresponding or similar products. Later, when the user lands on a website with an ad space, their browser executes locally the ad auction logic for all interest groups they belong to and display the ad for the winning bid. The advertiser with the winning bid would thus be able to show ads targeted based on users' browsing behavior.

- **API calls:**
    - Interest Groups:
        - `navigator.joinAdInterestGroup()`
        - `navigator.leaveAdInterestGroup()`
        - `navigator.clearOriginJoinedAdInterestGroups()`
    - Auctions:
        - `navigator.runAdAuction()`
        - `navigator.adAuctionComponents()`
        - `navigator.createAuctionNonce()`
    - Several new HTTP headers as well:
        - `Ad-Auction-Allowed`
        - `Ad-Auction-Only`
        - `Ad-Auction-Signals`
        - `Ad-Auction-Additional-Bid`
        - `X-fledge-bidding-signals-format-version`
        - `Data-Version`
        - `Sec-Ad-Auction-Fetch`
- Documentation (Web)
- Explainer
- Documentation (Android)

#### 4.13.3 Analyses

- [58] Mateusz Rumiński, Przemysław Iwańczak, and Łukasz Włodarczyk. 2022. Findings from the Early Fledge Experiments. https://doi.org/10.2139/ssrn.4219796
- [2] Mir Masood Ali, Binoy Chitale, Mohammad Ghasemisharif, Chris Kanich, Nick Nikiforakis, and Jason Polakis. 2023. Navigating Murky Waters: Automated Browser Feature Testing for Uncovering Tracking Vectors. In *Proceedings 2023 Network and Distributed System Security Symposium*, 2023. Internet Society, San Diego, CA, USA. https://doi.org/10.14722/ndss.2023.24072
- [74] Martin Thomson. 2024. Protected Audience Privacy Analysis. (March 2024). Retrieved from https://mozilla.github.io/ppa-docs/protected-audience.pdf
- [53] Michiel Philipse. 2024. Post-Third-Party Cookies: Analyzing Google's Protected Audience API. Master Thesis. Retrieved from https://www.cs.ru.nl/masters-theses/2024/M_Philipse___Post-Third-Party_Cookies_Analyzing_Google's_Protected_Audience_API..pdf
- [44] Minjun Long and David Evans. 2024. Evaluating Google's Protected Audience Protocol. *Proceedings on Privacy Enhancing Technologies* (2024). Retrieved from https://petsymposium.org/popets/2024/popets-2024-0147.php
- [15] Giuseppe Calderonio, Mir Masood Ali, and Jason Polakis. 2024. Fledging Will Continue Until Privacy Improves: Empirical Analysis of Google's Privacy-Preserving Targeted Advertising. In *33rd USENIX Security Symposium (USENIX Security 24)*, 2024. 4121–4138. Retrieved from https://www.usenix.org/conference/usenixsecurity24/presentation/calderonio

### 4.14 Related Website Sets

This API is deprecated, see the official announcement and this status overview from Google.

#### 4.14.1 Overview

Related Website Sets (RWS, previously named First Party Sets) is designed to minimize disruptions to specific user-facing features once Chrome starts limiting access to third-party cookies by default. The goal is to allow users to browse the web with minimal disruption while still upholding the privacy goals of the Privacy Sandbox. Specifically, RWS is a way for a company to declare relationships among sites, so that browsers allow limited third-party cookie access for specific purposes. Chrome will use these declared relationships to decide when to allow or deny a site access to their cookies when in a third-party context. At a high level, an RWS is a collection of domains, for which there is a single "set primary" and potentially multiple "set members."

#### 4.14.2 How does it work?

Check out our `.well-known` scanner and analysis code, more details also on this post.

A related website set consists of one primary site and up to five associated sites. To use a set, its JSON must be added to the `related_website_sets.JSON` file available on the RWS GitHub repository, which Chrome then consumes to get the list of sets to apply RWS behavior to. `.well-known` files. Each site in the set must also serve a `.well-known` file at `/.well-known/related-website-set.json`, which serves to verify the set structure and the relationship between the sites in the set. The primary site's `.well-known` file must explicitly list out the full set structure.

- Documentation
- MDN Documentation
- Explainer

The Storage Access API (SAA) provides a way for embedded cross-origin content to access the storage that it would normally only have access to in a first-party context. Embedded resources can use SAA methods to check whether they currently have access to storage, and to request access from the user agent. When third-party cookies are blocked but RWS is enabled, Chrome will automatically grant permission in intra-RWS contexts (such as an iframe, whose embedded site and top-level site are in the same RWS), and will show a prompt to the user otherwise.

#### 4.14.3 Analyses

- [68] Peter Snyder. 2022. First-Party Sets: Tearing Down Privacy Defenses Just as They're Being Built. Retrieved from https://brave.com/web-standards-at-brave/8-first-party-sets/
- [46] Stephen McQuistin, Peter Snyder, Hamed Haddadi, and Gareth Tyson. 2024. A First Look at Related Website Sets. In *Proceedings of the 2024 ACM on Internet Measurement Conference*, November 2024. 107–113. https://doi.org/10.1145/3646547.3689026
- [12] Yohan Beugin, Sam Dutton, Yana Dimova, Rowan Merewood, and Barry Pollard. 2024. The 2024 Web Almanac: Cookies. In *The 2024 Web Almanac*. HTTP Archive. https://doi.org/10.5281/zenodo.14065903

### 4.15 SDK Runtime

This API is deprecated, see the official announcement and this status overview from Google.

#### 4.15.1 Overview

The SDK (Software Development Kit) Runtime aimed to separate and isolate third-party code in Android applications from the main app's code to limit tracking.

- Documentation (Android)

#### 4.15.2 Analyses

- [45] Haoran Lu, Yichen Liu, Xiaojing Liao, and Luyi Xing. 2024. Towards Privacy-Preserving Social-Media SDKs on Android. In *33rd USENIX Security Symposium (USENIX Security 24)*, 2024. 647–664. Retrieved from https://www.usenix.org/conference/usenixsecurity24/presentation/lu-haoran

### 4.16 Shared Storage API

This API is deprecated, see the official announcement and this status overview from Google.

#### 4.16.1 Overview

To prevent cross-site tracking, browsers are partitioning access to cookies, web storage resources, IndexedDB, cache API, etc. However, some legitimate use cases rely on cross-site information sharing, such as when advertisers need to measure ad reach across sites, or publishers wants to customize user experience based on the interests, groups, or past interactions of the user. The Shared Storage API aims to provide the needed data storage, processing, and sharing solution for these use cases without enabling third-party tracking.

#### 4.16.2 How does it work?

This is done by allowing unlimited, cross-site storage write access with privacy-preserving read access. Indeed, Similar to other storage APIs, shared storage allows writing data at any time but restricts reading data to inside a *worklet*. Worklets offer a secure environment for processing shared storage data, although direct data sharing with the associated browsing context isn't permitted.

- **API call:**
  - The party can write into storage using `window.sharedStorage` API
  - Available commands : `set`,`append`,`delete`,`clear`
  - Read from shared storage : only allowed in a fenced frame initiated by the same party
- Documentation
- MDN Documentation
- Explainer

#### 4.16.3 Analyses

- [49] Alexandra Nisenoff, Deian Stefan, and Nicolas Christin. 2025. Exploiting the Shared Storage API. In *Proceedings of the 2025 ACM SIGSAC Conference on Computer and Communications Security (CCS '25)*, November 2025. Association for Computing Machinery, New York, NY, USA, 1260–1274. https://doi.org/10.1145/3719027.3744848

### 4.17 Storage and Network State Partitioning

This API is still being maintained by Google in Chrome, even after the deprecation announcement of most other Privacy Sandbox APIs.

#### 4.17.1 Overview

Like other major web browsers such as Firefox and Safari, Chrome is preventing certain side-channel cross-site tracking by isolating storage and network communication APIs in third-party context.

- Documentation
- Storage Partition Explainer
- Network State Partition Explainer

### 4.18 Topics API

This API is deprecated, see the official announcement and this status overview from Google.

#### 4.18.1 Overview

**On the Web:** The Topics API is being developed by Google to deprecate *third-party cookies* on Chrome with the promise of both providing *utility to advertisers* and *privacy to users*. Google claims that topics do not reveal sensitive information because they are picked from a curated taxonomy, that users have *plausible deniability* as noisy topics are sometimes returned by the API, and that Topics can not be used to fingerprint users as others share the same interests (*k-anonymity*). According to Google, observed topics are *useful* to advertisers because they are aligned with users' interests and the API is designed so that it can be *usable and understandable* by users.

**On Android:** The goal of the Topics API for Android is to enable interest-based advertising on Android while deprecating the use of cross-app identifiers such as the Android advertising ID. Google expects that application developers and third parties will not be able to track users across applications, or even need to do so if the returned topics are aligned well enough with what advertisers want to infer about users' interests. Google claims that third parties can not infer which applications are installed on users' devices from the topics that are returned.

#### 4.18.2 How does it work?

Check out our exact reimplementation of the Topics Classifier (Web + Android), more details also on this post.

**On the Web:** The Topics API works by having the web browser classify the websites visited by users into categories of interest. Advertisers who are embedded on websites can observe some of the recent top users' topics and use that information to perform an ad auction.

In practice, the latest taxonomy is composed of 469 topics that are each associated with a standard or high utility label. Hostnames of websites visited by users are classified by Chrome locally into topics using a list of about 50k annotated domains and an ML classifier. Each week and for each user, only the top 5 topics ordered by utility first and then frequency are saved.

When an advertiser makes a call to the Topics API, a maximum of 3 topics (1 for each of the last 3 weeks) are returned; 95% of the time a real topic is picked from the user's top 5 for that week, in the remaining 5%, a noisy one is randomly drawn from the taxonomy. Finally, a witness requirement in Topics enforces that a real topic is returned to an API caller only if they were embedded on a website of the same topic that was also visited by the user during one of the past 3 weeks.

- **API:**
  - `document.browsingTopics()` (can be passed optional argument `{skipObservation:true}` to not participate in topics observation)(+ also through HTTP Headers) returns promise to an array of up to 3 topics.
  - `Sec-Browsing-Topics` header of a `fetch()` request, `Observe-Browsing-Topics: ?1` header should be sent in the response to the request to participate in topics observation.
  - Opt-out for websites:
    - `Permissions-Policy: browsing-topics=()` on each page Topics API needs to be blocked
    - `Permissions-Policy: browsing-topics=(self "https://example.com")` to control who can call the API
- Documentation (Web)
- Explainer (Web)
- See `chrome://topics-internals` to see info about model version, taxonomy used, etc.

**On Android:** Similarly to the Topics API for the web classifying users' web behaviors into categories, the Topics API for Android monitors the applications being opened and used every week and classify users' behaviors into topics of interest. Then, Topics reports some of each user's top topics to third party SDKs that are embedded into other applications when they call the API.

- **API call:** `getTopics()`
- Documentation (Android)
- Documentation (Android)
- Explainer (Android)

**On the Web:** Google released a few privacy analyses about the re-identification risk of the Topics API. However, the empirical measurements are always done on a private dataset of real histories and only aggregate results reported, which prevents verification of Google's assertions. Analyses by other actors have pointed to the risk that Topics be used to fingerprint users across sites, and this was demonstrated to be the case on real and simulated datasets.

**On Android:** The Topics API for Android appeared to be the Web proposal directly taken and adapted to Android applications: i.e., web domain names replaced by application identifiers. Thus, the Android proposal likely inherits similar limitations than the Topics API for the Web, but it is also unclear if and how the specifics of the Android ecosystem might be different and impact the API guarantees and users' privacy. For instance: - Would users browsing the web on Chrome on their Android devices have both "Web" and "Android" topics profiles and would applications (like Chrome) be able to access both? Could profiles be linked? - Advertisers also need to provide a public key used to encrypt the topics returned to them on Android, while the Web implementation does not require that. The reasoning behind this extra encryption step on Android is a bit unclear: is the concern that other SDKs would observe the results of API calls triggered by and aimed for a different APK?

#### 4.18.3 Analyses

- [67] Peter Snyder. 2022. Google's Topics API: Rebranding FLoC Without Addressing Key Privacy Issues. Retrieved from https://brave.com/web-standards-at-brave/7-googles-topics-api/
- [24] Alessandro Epasto, Andres Munoz Medina, Christina Ilvento, and Josh Karlin. 2022. Measures of Cross-Site Re-Identification Risk: An Analysis of the Topics API Proposal. (2022), 12. Retrieved from https://github.com/patcg-individual-drafts/topics/blob/main/topics_analysis.pdf
- [40] Anne van Kesteren. 2023. WebKit Standards Positions - The Topics API. Retrieved from https://github.com/WebKit/standards-positions/issues/111
- [16] C. J. Carey, Travis Dick, Alessandro Epasto, Adel Javanmard, Josh Karlin, Shankar Kumar, Andres Munoz Medina, Vahab Mirrokni, Gabriel Henrique Nunes, Sergei Vassilvitskii, and Peilin Zhong. 2023. Measuring Re-Identification Risk. https://doi.org/10.48550/arXiv.2304.07210
- [73] Martin Thomson. 2023. A Privacy Analysis of Google's Topics Proposal. (January 2023). Retrieved from https://mozilla.github.io/ppa-docs/topics.pdf
- [35] Nikhil Jha, Martino Trevisan, Emilio Leonardi, and Marco Mellia. 2023. On the Robustness of Topics API to a Re-Identification Attack. *Proceedings on Privacy Enhancing Technologies* (2023). Retrieved from https://petsymposium.org/popets/2023/popets-2023-0098.php
- [36] Nikhil Jha, Martino Trevisan, Emilio Leonardi, and Marco Mellia. 2024. Re-Identification Attacks against the Topics API. *ACM Trans. Web* 18, (August 2024), 39:1–39:24. https://doi.org/10.1145/3675400
- [3] Mário S. Alvim, Natasha Fernandes, Annabelle McIver, and Gabriel H. Nunes. 2023. A Quantitative Information Flow Analysis of the Topics API. In *Proceedings of the 22nd Workshop on Privacy in the Electronic Society (WPES '23)*, November 2023. Association for Computing Machinery, New York, NY, USA, 123–127. https://doi.org/10.1145/3603216.3624959
- [4] Mário S. Alvim, Natasha Fernandes, Annabelle McIver, and Gabriel H. Nunes. 2024. The Privacy-Utility Tradeoff in the Topics API. In *Proceedings of the 2024 on ACM SIGSAC Conference on Computer and Communications Security*, December 2024. 1106–1120. https://doi.org/10.1145/3658644.3670368
- [78] Alberto Verna, Nikhil Jha, Martino Trevisan, and Marco Mellia. 2024. A First View of Topics API Usage in the Wild. In *Proceedings of the 20th International Conference on Emerging Networking EXperiments and Technologies (CoNEXT '24)*, December 2024. Association for Computing Machinery, New York, NY, USA, 48–54. https://doi.org/10.1145/3680121.3697810
- [79] Alberto Verna, Nikhil Jha, Martino Trevisan, and Marco Mellia. 2025. Understanding Topics API in the Wild: Dubious Usage and Stale Adoption. *IEEE Transactions on Privacy* 2, (2025), 119–130. https://doi.org/10.1109/TP.2025.3615120
- [8] Yohan Beugin and Patrick McDaniel. 2024. A Public and Reproducible Assessment of the Topics API on Real Data. In *IEEE Security and Privacy Workshop on Designing Security for the Web (SecWeb)*, May 2024. https://doi.org/10.48550/arXiv.2403.19577
- [10] Yohan Beugin and Patrick McDaniel. 2024. Interest-disclosing Mechanisms for Advertising are Privacy-Exposing (not Preserving). In *Proceedings on Privacy Enhancing Technologies Symposium (PETS)*, July 2024. https://doi.org/10.56553/popets-2024-0004
- [22] Travis Dick, Alessandro Epasto, Adel Javanmard, Josh Karlin, Andrés Muñoz Medina, Vahab Mirrokni, Sergei Vassilvitskii, and Peilin Zhong. 2025. Differentially Private Synthetic Data Release for Topics API Outputs. In *Proceedings of the 31st ACM SIGKDD Conference on Knowledge Discovery and Data Mining V.2 (KDD '25)*, August 2025. Association for Computing Machinery,

New York, NY, USA, 5379–5390. https://doi.org/10.1145/3711896.3737391
- [70] Athicha Srivirote, Muhammad Abu Bakar Aziz, Jeffrey Gleason, Desheng Hu, and Christo Wilson. 2026. Inferring Users' Demographics and Sensitive Interests Using the Topics API. In *Proceedings of the ACM Web Conference 2026 (WWW '26)*, April 2026. Association for Computing Machinery, New York, NY, USA, 3438–3448. https://doi.org/10.1145/3774904.3792660

### 4.19 User-Agent Reduction & User-Agent Client Hints

This API is still being maintained by Google in Chrome, even after the deprecation announcement of most other Privacy Sandbox APIs.

#### 4.19.1 Overview

User-Agent reduction is a mechanism implemented by several web browsers to minimize the amount of identifying information (used by fingerprinting techniques) shared through the `User-Agent` header and other APIs such as `navigator.userAgent`, `navigator.appVersion`, and `navigator.platform`.

User-Agent Client Hints allow web servers to explicitly request access to features (also designated as hints by the proposal) of the User-Agent header that are not exposed by default anymore. The idea is that browsers share a user-agent with low-entropy hints during the initial request, then servers that need them can explicitly ask compatible browsers for more hints through the proposed `Accept-CH` header. For a more detailed example, see this explanation.

Browsers would thus be technically able to mediate access to these user-agent hints of higher entropy. However, the proposal is unclear on how to perform so in practice, moreover, Google decided to not pursue their privacy budget proposal efforts that may have been part of a solution to this mediation problem.

- Documentation User-Agent Reduction
- Documentation User-Agent Client Hints

#### 4.19.2 Analyses

- [64] Peter Snyder, Pranjal Jumde, Tom Lowenthal, and Brian Clifton. 2019. Brave's Concerns with the Client-Hints Proposal. Retrieved from https://brave.com/web-standards-at-brave/1-client-hints/
- [33] Jean Luc Intumwayase, Imane Fouad, Pierre Laperdrix, and Romain Rouvoy. 2023. UA-Radar: Exploring the Impact of User Agents on the Web. In *Proceedings of the 22nd Workshop on Privacy in the Electronic Society (WPES '23)*, November 2023. Association for Computing Machinery, New York, NY, USA, 31–43. https://doi.org/10.1145/3603216.3624958
- [60] Asuman Senol and Gunes Acar. 2023. Unveiling the Impact of User-Agent Reduction and Client Hints: A Measurement Study. In *Proceedings of the 22nd Workshop on Privacy in the Electronic Society (WPES '23)*, November 2023. Association for Computing Machinery, New York, NY, USA, 91–106. https://doi.org/10.1145/3603216.3624965
- [81] Stephan Wiefling, Marian Hönscheid, and Luigi Lo Iacono. 2024. A Privacy Measure Turned Upside Down? Investigating the Use of HTTP Client Hints on the Web. In *Proceedings of the 19th International Conference on Availability, Reliability and Security (ARES '24)*, July 2024. Association for Computing Machinery, New York, NY, USA, 1–12. https://doi.org/10.1145/3664476.3664478
- [32] Sarp Ilgaz. 2025. Investigating High Entropy Client Hint usage in HTTP/2 and HTTP/3. Bachelor Thesis. Retrieved from https://www.cs.ru.nl/bachelors-theses/2025/Sarp_Ilgaz___1093588___Investigating_High_Entropy_Client_Hint_usage_in_HTTP-2_and_HTTP-3.pdf

## 5 Other Proposals & Mechanisms

Below is a list of other proposals and mechanisms from Google or other actors that are also related to online advertising, security, and privacy along with their corresponding analyses. See this overview for the Privacy Sandbox proposals.

### 5.1 DNS-over-HTTPS

This API has been launched and is now a standard: RFC 8484.

#### 5.1.1 Overview

DNS-over-HTTPS is a protocol that encrypts Domain Name System (DNS) queries and responses by encoding them within HTTPS messages. This helps prevent attackers from sending users to phishing websites or observing what sites they visit.

- Blog Post
- Standard

#### 5.1.2 Analyses

- [59] Shivan Kaul Sahib and Peter Snyder. 2021. Encrypting DNS Zone Transfers. Retrieved from https://brave.com/web-standards-at-brave/5-encrypting-dns-zone-transfers/

### 5.2 Global Privacy Control

Global Privacy Control (GPC) has been implemented by several browsers and extensions. GPC is a valid do-not-sell-my-personal-information signal that businesses, as of November 2025, are legally required to respect by the legislation of four U.S. states (California, Colorado, Connecticut, and New Jersey).

- **Proposed by** Ashkan Soltani (Georgetown Law) and Sebastian Zimmeck (Wesleyan University) in collaboration with The New York Times, The Washington Post, Financial Times, Automattic (Wordpress.com & Tumblr), Glitch, DuckDuckGo, Brave, Mozilla, Disconnect, Abine, Digital Content Next (DCN), Consumer Reports, and the Electronic Frontier Foundation (EFF).
- Website
- Specification

#### 5.2.1 Analyses

- [62] Peter Snyder and Anton Lazarev. 2020. Global Privacy Control, a New Privacy Standard Proposal. Retrieved from https://brave.com/web-standards-at-brave/4-global-privacy-control/

### 5.3 Interoperable Private Attribution

- **Proposed by:** Erik Taubeneck (Meta), Ben Savage (Meta), Martin Thomson (Mozilla)
- Explainer
- Implementation
- Blog post

#### 5.3.1 Analyses

- [17] Benjamin Case, Richa Jain, Alex Koshelev, Andy Leiserson, Daniel Masny, Thurston Sandberg, Ben Savage, Erik Taubeneck, Martin Thomson, and Taiki Yamaguchi. 2023. Interoperable Private Attribution: A Distributed Attribution and Aggregation Protocol. Retrieved from https://eprint.iacr.org/2023/437

### 5.4 Privacy Pass

Privacy Pass is available as a browser extension, has been implemented by Cloudflare, Apple, and others, and is an official IETF standard (RFC 9576, RFC 9577, RFC 9578).

- **Proposed by** Alex Davidson (Royal Holloway, University of London (work completed during an internship at Cloudflare)), Ian Goldberg (University of Waterloo), Nick Sullivan (Cloudflare), George Tankersley, Filippo Valsorda
- Initial Website
- IETF Working Group
- Cloudflare Research Website

#### 5.4.1 Analyses

- [20] Alex Davidson, Ian Goldberg, Nick Sullivan, George Tankersley, and Filippo Valsorda. 2018. Privacy Pass: Bypassing Internet Challenges Anonymously. *Proceedings on Privacy Enhancing Technologies* (2018). Retrieved from https://petsymposium.org/popets/2018/popets-2018-0026.php
- [29] Konrad Hanff, Anja Lehmann, and Cavit Özbay. 2025. Security Analysis of Privately Verifiable Privacy Pass. In *Proceedings of the 2025 ACM SIGSAC Conference on Computer and Communications Security (CCS '25)*, November 2025. Association for Computing Machinery, New York, NY, USA, 2922–2936. https://doi.org/10.1145/3719027.3765172

### 5.5 Private Click Measurement

- **Proposed by:** Apple
- Explainer
- Blog post

#### 5.5.1 Analyses

- [72] Martin Thomson. 2022. An Analysis of Apple's Private Click Measurement. (June 2022). Retrieved from https://mozilla.github.io/ppa-docs/pcm.pdf

### 5.6 SWAN & Unified ID 2.0

SWAN and Unified ID 2.0 are two proposals from different actors but with some similarities; they both propose to assign users a pseudonymous identifier that can be used for cross-site tracking and would not be stored in third-party cookies. SWAN would do so through bounce tracking, while Unified ID 2.0 would use identifiers such as email addresses.

#### 5.6.1 SWAN

- **Proposed by:** 51Degrees
- Explainer

#### 5.6.2 Unified 2.0

- **Proposed by:** IAB Tech Lab
- Explainer
- Website

#### 5.6.3 Analyses

- [71] Martin Thomson and Eric Rescorla. 2021. Comments on SWAN and Unified ID 2.0. (August 2021). Retrieved from https://mozilla.github.io/ppa-docs/swan_uid2_report.pdf

### 5.7 Tech Lab Trusted Server Initiative

The IAB Tech lab has launched in November 2025, the Trusted Server Initiative that proposes moving advertising third-party code to the publisher's server side (i.e., first party).

### 5.8 TURTLEDOVE/PARAKEET Proposals

Chrome created the original TURTLEDOVE proposal that led to the following series of somewhat related proposals and mechanisms from different actors.

#### 5.8.1 TURTLEDOVE (original)

- **Proposed by:** Google Chrome
- Explainer

#### 5.8.2 SPARROW

- **Proposed by:** Criteo
- Explainer

#### 5.8.3 PARAKEET

- **Proposed by:** Microsoft
- Explainer

#### 5.8.4 Outcome-based and Product-level TURTLEDOVE

- **Proposed by:** RTB House
- Outcome-based Explainer
- Product-level Explainer

#### 5.8.5 Dovekey

- **Proposed by:** Google Ads
- Explainer

#### 5.8.6 PARROT

- **Proposed by:** Magnite
- Explainer

#### 5.8.7 TERN

- **Proposed by:** NextRoll
- Explainer

#### 5.8.8 Frequency Capping

- **Proposed by:** Google
- Proposal Explainer
- Experiment Explainer

#### 5.8.9 And other proposals from Google

- **Proposed by:** Google
- List of proposals
- List of experiments

#### 5.8.10 Analyses

- [34] Przemysław Iwańczak and Mateusz Ruminski. 2022. The Future of Frequency Capping in Privacy-Centric Digital Advertising. https://doi.org/10.2139/ssrn.3985974

### 5.9 Web Bundles

This API is deprecated (the experimentation was removed from Chrome in February 2023).

- **Proposed by:** Google
- Explainer
- Documentation

#### 5.9.1 Analyses

- [47] Mozilla. 2019. Mozilla Position on Web Packaging. Retrieved from https://docs.google.com/document/d/1ha00dSGKmjoEh2mRiG8FIA5sJ1KihTuZe-AXX1r8P-8/edit?usp=sharing
- [65] Peter Snyder. 2020. WebBundles Harmful to Content Blocking, Security Tools, and the Open Web. Retrieved from https://brave.com/web-standards-at-brave/3-web-bundles/

### 5.10 Web Environment Integrity

This proposal is no longer pursued, but an Android API is being considered.

- **Proposed by:** Google
- Explainer

#### 5.10.1 Analyses

- [69] Peter Snyder. 2023. ``Web Environment Integrity'': Locking Down the Web. Retrieved from https://brave.com/web-standards-at-brave/9-web-environment-integrity/
- [31] Cory Doctorow and Jacob Hoffman-Andrews. 2023. Your Computer Should Say What You Tell It To Say. Retrieved from https://www.eff.org/deeplinks/2023/08/your-computer-should-say-what-you-tell-it-say-1
- [14] Grinstead Brian. 2023. Request for Mozilla Position on an Emerging Web Specification. Retrieved from https://github.com/mozilla/standards-positions/issues/852
- [54] Julien Picalausa. 2023. Unpacking Google's New ``Dangerous'' Web-Environment-Integrity Specification. Retrieved from https://vivaldi.com/blog/googles-new-dangerous-web-environment-integrity-spec/
- [30] Philippe Le Hégaret. 2023. Web Environment Integrity Has No Standing at W3C; Understanding New W3C Work. Retrieved from https://www.w3.org/blog/2023/web-environment-integrity-has-no-standing-at-w3c/

## Acknowledgements

This material is based upon work supported by the National Science Foundation under Grant No. CNS-1900873, CNS-232088, and CNS-2343611. Any opinions, findings, and conclusions or recommendations expressed in this material are those of the author(s) and do not necessarily reflect the views of the National Science Foundation. This work was also supported in part by the Semiconductor Research Corporation (SRC) and DARPA.